\documentclass[aps, prl, preprint, graphics, showpacs, amssymb, amsmath, superscriptaddress]{revtex4-1}
\usepackage{graphicx}
\usepackage{dcolumn}
\usepackage{bm}
\usepackage{epstopdf}
\usepackage{amssymb}
\usepackage{color}
\usepackage[colorlinks=true, letterpaper=true, pdfstartview=FitV, linkcolor=blue, citecolor=blue, urlcolor=blue]{hyperref}
\usepackage{lineno}

\newcommand{\busedit}[1]{\textbf{\textcolor{red}{#1}}}
\usepackage{soul}

\begin{document}

\title{Emergent Multifunctionality in Two-Dimensional Janus VSBrI Monolayer: A Study of Multiferroicity, Magnetoelectricity, and Piezoelectricity}
\author{Qiuyue Ma}
\author{Busheng Wang}
\author{Guochun Yang}
\author{Yong Liu}\email{yongliu@ysu.edu.cn}
\affiliation{State Key Laboratory of Metastable Materials Science and Technology \& Key Laboratory for Microstructural Material Physics of Hebei Province, School of Science, Yanshan University, Qinhuangdao 066004, China}

\date{\today}

\begin{abstract}
The Janus VSBrI monolayer, identified by first-principles calculations, emerges as a promising semiconducting material with desirable ferromagnetic and ferroelectric properties. It's in-plane magnetic anisotropic energy is up to 460 $\mu$eV/V , and in-plane and out-of-plane piezoelectric strain coefficients are larger than many other known two-dimensional materials. The energy variations among different magnetic states show a strong correlate with polarization. Interestingly, the stability of the ferroelectric phase can be further enhanced by the application of biaxial tensile strain. These intriguing properties make the Janus VSBrI monolayer highly desirable for practical applications in piezoelectronic devices, and a promising candidate for  multifunctional spintronic devices.

\end{abstract}

\maketitle


\maketitle
\busedit{\st{\section{INTRODUCTION}}}

The discovery of two-dimensional (2D) materials, characterized by their atomic-scale thickness, has revolutionized fundamental physics research, especially following the isolation of graphene~\cite{1K-Science-2004,3K-Nature-2005,4E-Phys.-2007}. These materials offer exceptional physical properties and enhance device performance, owing to their multifunctional nature and ultra-thin structure, thus opening new possibilities in fields like optoelectronics and spintronics~\cite{H-Nat.-2018,C-Natl-2019}. However, achieving multifunctionality in 2D materials remains a challenge due to the complex interactions among charge, spin, and lattice structures~\cite{W-Nature-2006,S-Adv-2015}. Multiferroic materials, which exhibit more than one ferroic ordering (e.g., ferroelectricity, ferromagnetism, ferroelasticity), have garnered attention, especially with recent advancements in 2D ferroelectric and ferromagnetic materials~\cite{C.H-Phys-2018,S.-Nat-2007}. Despite their potential, 2D magnetoelectric multiferroics, which couple ferromagnetic and ferroelectric properties, remain scarce and difficult to design~\cite{N-Science-2005}. This challenge arises from the inherent conflict between the requirements for magnetism and ferroelectricity, particularly highlighted by the ``\emph{d$_{0}$} rule" in perovskites~\cite{N-J. Phys. Chem. B-2000}. While magnetism requires partially filled \emph{d} orbitals, ferroelectricity demands empty \emph{d} orbitals, creating a fundamental contradiction that limits the development of single-phase magnetoelectric materials. Nonetheless, recent research efforts~\cite{K.-Adv. Phys.-2009,J.-Nature-2010} have focused on overcoming these challenges to better understand and design magnetoelectric multiferroics for advanced applications.


Janus monolayers, a novel class of 2D materials with mirror asymmetry, have generated significant attention across various fields, from separation membranes to advanced 2D electronics. Their distinctive vertical asymmetry gives rise to remarkable physical phenomena, such as the Rashba effect~\cite{T.-Phys.-2018}, valley spin splitting~\cite{J.-Phys. -2024,S.-Phys. -2020}, catalytic performance~\cite{D.-Nano-2018,W.-J.-2018}, and piezoelectric polarization~\cite{L.-ACS-2017}. Transition metal dichalcogenide (TMD) Janus monolayers, hold significant potential in spintronics, photocatalysis, and valleytronics applications. For example, the Janus WSeTe monolayer shows large Rashba spin splitting~\cite{Q.-F-Phys.-2017}, Janus MoSSe has demonstrated excellent photocatalytic properties for solar-driven water splitting\cite{X.-J. Mater-2018}, and Janus VSSe exhibits promising piezoelectricity, ferroelasticity, and valley polarization~\cite{CZhang-Nano-2019}. These exceptional properties position Janus monolayers as key materials for the future of electronic and spintronic devices.

The piezoelectric effect, a key electromechanical phenomenon, occurs in semiconductors with non-centrosymmetric crystal structures~\cite{W.-Nat.-2016,C.-J. Am-2020} and is especially prevalent in Janus materials. The successful synthesis of the Janus MoSSe monolayer, which exhibits vertical dipoles and piezoelectric properties, highlights its potential for future technological applications~\cite{A.-Nat.-2017}. Theoretically, other 2D Janus materials such as group-III chalcogenide monolayers~\cite{Y-Appl.-2017}, ZnBrI~\cite{L.-Chem.-2022}, FeGeN$_3$~\cite{Z. -Appl.-2024}, and Bi$_2$X$_2$Y (X = S, Se, Te; Y = S, Se, Te; X $\neq$ Y)~\cite{J.-Mater.-2021} have also been explored for their piezoelectric characteristics, expanding the scope of 2D materials in this field. However, a critical challenge remains: the absence or weak manifestation of out-of-plane piezoelectricity in many 2D systems. Consequently, the quest for materials with strong out-of-plane piezoelectric responses has become a pivotal area of research.

In this letter, we introduce a Janus VSBrI monolayer as a multifunctional material based on first-principles calculations. It exhibits intrinsic ferromagnetic, ferroelectric, and piezoelectric properties, positioning it as a promising candidate for advanced electronic applications. It is a semiconductor with a bandgap of 0.76 eV, displaying strong magnetic anisotropy (460 $\mu$eV/V) and an in-plane magnetic axis, with a Curie temperature estimated at 83 K. The Janus VSBrI monolayer also demonstrates spontaneous in-plane ferroelectric polarization of 1.20 $\times$ $10^{-10}$ C/m, significant magnetoelectric coupling, and notable piezoelectric nature. These combined characteristics make Janus VSBrI monolayer a highly attractive platform for future multiferroic and spintronics devices.


\busedit{\st{\section{Methods}}}
The structural optimization and properties calculations are performed within the framework of density functional theory using a planewave basis set as implemented in Vienna $ab$ $initio$ Simulation Package (VASP)~\cite{G. Kresse-Computational Materials Science-1996,G-Phys. Rev. B-1996}. The projector-augmented wave (PAW) potential with the generalized gradient approximation of Perdew-Burke-Ernzerhof (GGA-PBE) is used as the exchange-correlation functional~\cite{J. P-Phys. Rev. Lett.-1996}. Considering the strong electronic correlation of 3\emph{d} electrons, an effective Hubbard $U$ ($U_{eff}$ = 2 eV) is added for the V-3\emph{d} orbitals~\cite{X.-Nanoscale-2022}. The plane wave cutoff energy is set to 450 eV, and a $\Gamma$-centered 15 $\times$ 13 $\times$ 1 Monkhorst-Pack grid is employed to sample the first Brillouin zone. A vacuum layer of at least 15 {\AA} along the \emph{z} direction is added to avoid unexpected interactions between periodic images. The structure undergoes full optimization until the Hellmann-Feynman forces acting on each atom are below 0.01 eV/{\AA}, while ensuring that the total energy difference between consecutive steps remains less than 10$^{-6}$ eV. The stability dependence on temperature of 4 $\times$ 4 $\times$ 1 supercell is confirmed through $ab$ $initio$ molecular dynamics (AIMD) simulations~\cite{v-J. Chem. Phys.-1992}. Phonon spectrum calculations are self-consistently conducted using density functional perturbation theory (DFPT) \cite{S.-Rev. Mod. Phys.-2001} implemented in the PHONOPY code \cite{A-Phys. Rev. B-2008}. The ferroelectric polarization is calculated by using the standard Berry phase method~\cite{R-Phys. Rev. B-1993}. The elastic stiffness tensor \emph{C$_{ij}$} and piezoelectric stress tensor \emph{e$_{ij}$} are calculated by using the strain-stress relationship (SSR) and DFPT method, respectively.

\busedit{\st{\section{RESULTS and DISCUSSION}}}
\begin{figure}[hb!]
\centerline{\includegraphics[width=0.85\textwidth]{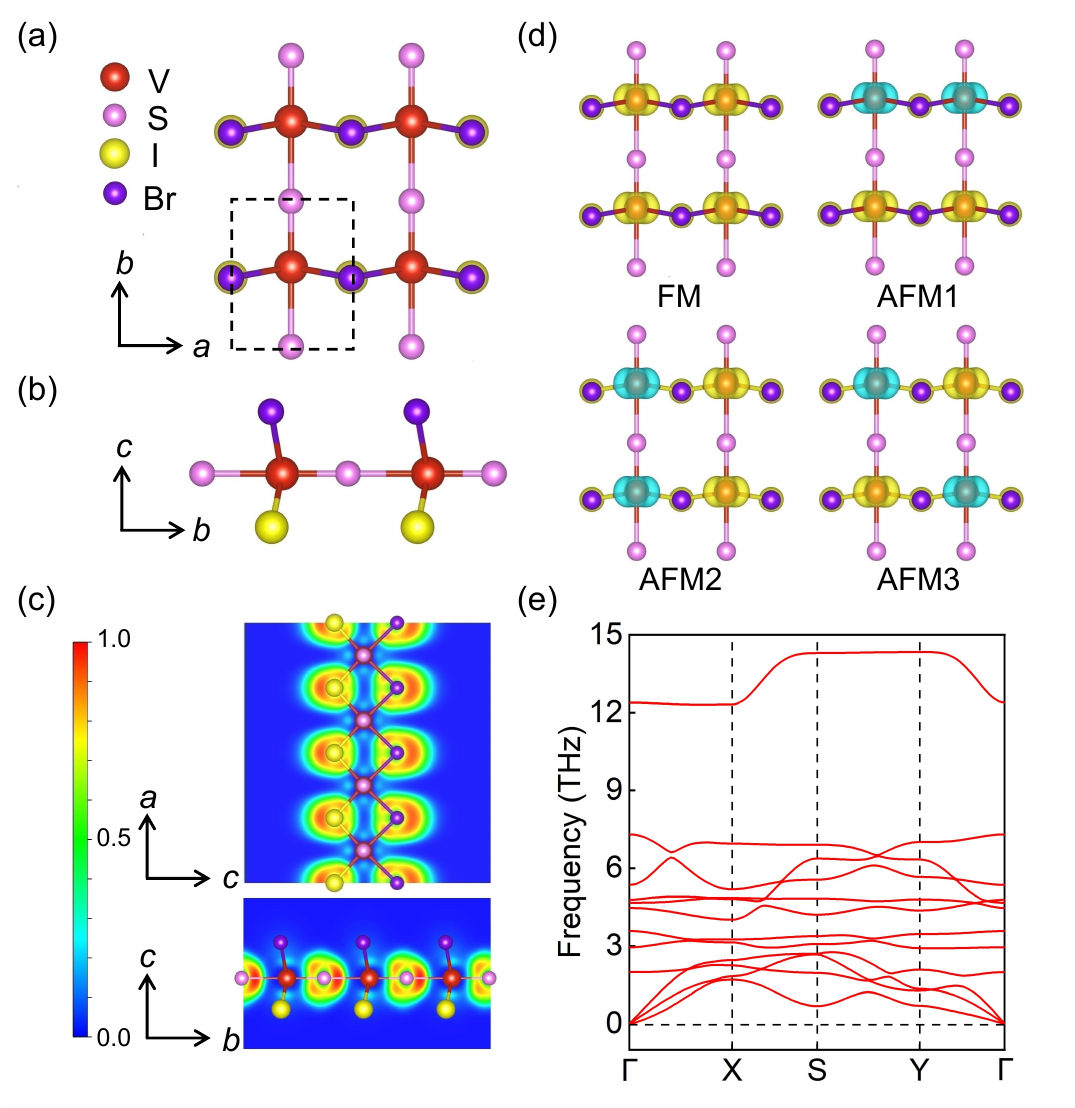}}
\caption{(a) Top and (b) side views of the Janus VSBrI monolayer. The unit cell is marked by the black dashed lines. (c) Electron localization function (ELF) plot of the Janus VSBrI monolayer along the 010 and 100 plane. (d) The spin density isosurfaces of the Janus VSBrI monolayer for the four magnetic configurations. Yellow and green isosurfaces correspond to positive and negative spin densities, respectively. (e) The phonon dispersion of the Janus VSBrI monolayer.
\label{fig:1}}
\end{figure}

Figures~\ref{fig:1}(a) and (b) show the top and side views of the Janus VSBrI monolayer, which is derived from the ground-state structure of the parent VSI$_2$ monolayer~\cite{D.-Appl.-2023} by replacing the top layer of halogen atoms I with the more electronegative halogen Br. The Janus nature of VSBrI arises from the presence of two outer sublayers consisting of nonequivalent halogen atoms sandwiching the central V atoms, resulting in a lower symmetry (\emph{Pm}) in comparison to its parent VSI$_2$ monolayer (\emph{Pmm2}). As shown in Fig.~\ref{fig:1}(c), the electron localization function (ELF) illustrates the ionic characteristics of V-S, V-I, and V-Br bonds, and bader charge analysis indicates that S, Br, and I atoms gain electrons, while V atoms lose electrons. Four magnetic configurations of V atoms (one FM state and three AFM states) with the supercell of 2 $\times$ 2 $\times$ 1 are considered to determine the magnetic ground state of the Janus VSBrI monolayer [Fig.~\ref{fig:1}(d)]. By comparing the energies of different magnetic configurations, it is determined that the magnetic ground state of the Janus VSBrI monolayer is FM, with energies of 23.18, 25.67, and 19.25 meV/f.u. relative to the AFM1, AFM2, and AFM3 states, respectively. Spatial distribution of the spin-polarized electron density reveals that the significant FM localized magnetic moment (1 $\mu_B$ per unit cell) is primarily contributed by the V atom. The optimized lattice parameters for the FM unit cell are $a$ = 3.85 {\AA} and $b$ = 4.66 {\AA} (Structural parameters are given in Table SI of the Supplemental Material~\cite{Suppl.Mata.}).

To assess the dynamical stability, we conducted phonon dispersion spectrum calculations for the Janus VSBrI monolayer, illustrated in Fig.~\ref{fig:1}(e). The absence of imaginary frequencies confirms its dynamic stability. Its good thermal stability is evidenced by the little energy fluctuation and  slight structural deformation observed in the AIMD simulation conducted at a constant room temperature of 300 K [Fig. S1]~\cite{Suppl.Mata.}. A cohesive energy value of 3.46 eV/atom for the Janus VSBrI monolayer demonstrates a level of stability comparable to that of phosphorene (3.48 eV/atom)~\cite{J.-Phys.-2014}. The calculated elastic constants (\emph{C$_{11}$} = 37.45 N/m , \emph{C$_{12}$} = 7.49 N/m, \emph{C$_{22}$} = 31.48 N/m, and \emph{C$_{66}$} = 9.96 N/m) satisfy the Born-Huang criteria for mechanical stability: \emph{C$_{11}$} $>$ 0, \emph{C$_{66}$} $>$ 0, \emph{C$_{11}$}\emph{C$_{22}$} - $\emph{C$_{12}$}^2$ $>$ 0, indicating that the Janus VSBrI monolayer is mechanically stable.

\begin{figure}[t]
\centerline{\includegraphics[width=0.9\textwidth]{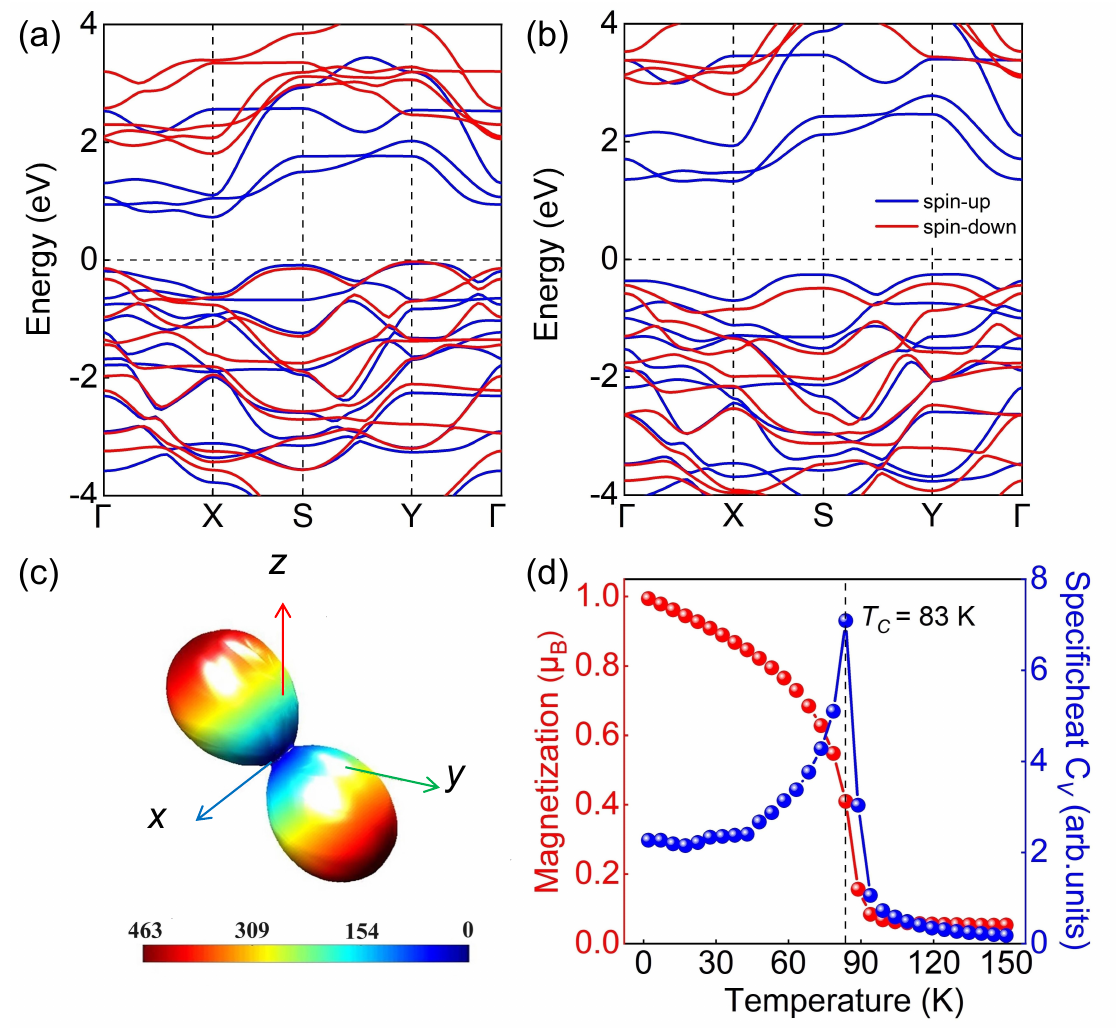}}
\caption{The calculated band structures of the Janus VSBrI monolayer by using (a) PBE + $U$, and (b) HSE06 functional. (c) Angular dependence of the magnetic anisotropy energy (MAE) of the Janus VSBrI monolayer with the magnetization direction lying in three-dimensional space. (d) Magnetic moments and specific heat as a function of temperature for the Janus VSBrI monolayer.
\label{fig:2}}
\end{figure}

As illustrated in Fig.~\ref{fig:2}(a), the VSBrI monolayer is identified as an indirect semiconductor with a band gap of 0.76 eV, as calculated using the revised PBE + $U$ functional. Its band gap value is larger than that of the parent VSI$_2$ monolayer (0.63 eV)~\cite{D.-Appl.-2023}. It is important to note that the hybrid Heyd-Scuseria-Ernzerh (HSE06) functional exhibits similar band structure to the PBE + $U$ functional. However, the values of band gaps (1.58 eV in spin-up, and 3.21 eV in spin-down) are larger than those (0.78 eV in spin-up, and 1.83 eV in spin-down) of the PBE + $U$ functional [Fig.~\ref{fig:2}(b)]. The density of states (DOS) of the Janus VSBrI monolayer is shown in Fig. S2~\cite{Suppl.Mata.}, which indicates that the S and I ions contribute to the valence band maximum, while the conduction band minimum is primarily influenced by the V and S ions. Additionally, near the Fermi level, the spin-up states are  primarily contributed by the electronic states of V and I ions.

Magnetic anisotropy energy (MAE) can resist disruptions to magnetic ordering caused by external magnetic fields or thermal agitation. The relative energy associated with different magnetization directions indicates that the Janus VSBrI monolayer tends to favor in-plane magnetization (\emph{x}-axis) [Fig.~\ref{fig:2}(c)]. Its calculated MAE 460 $\mu$eV/V, is comparable to that of Mn$_2$PSb monolayer (450 $\mu$eV/Mn)~\cite{Q.-Appl.-2022} and larger than those of Cr$_{1.5}$Br$_{1.5}$I (356 $\mu$eV/Cr)~\cite{S.-Appl.-2021}, VP (101.4 $\mu$eV/V), and VAs (262.6 $\mu$eV/V)~\cite{X.-Phys.-2021} monolayers. Then, we determined the magnetic coupling parameters (\emph{J}) for the Janus VSBrI monolayer using the classical Heisenberg model. The Hamiltonian is defined as:

\begin{align}
H = - \sum\limits_{\langle i,j \rangle} J_{ij} S_i S_j - A (S_i^Z)^2. \tag{1}
\end{align}

where the \emph{J} includes the nearest-neighbor \emph{J$_1$}, next-nearest-neighbor \emph{J$_2$}, and third-nearest-neighbor \emph{J$_3$} (More details are given in the Supplemental Material~\cite{Suppl.Mata.}). The calculated \emph{J$_1$}, \emph{J$_2$}, and \emph{J$_3$} are 21.77, 16.79 and 14.82 meV, respectively. With the above-mentioned MAE and magnetic coupling parameters, Curie temperature (\emph{T}$_C$) of the Janus VSBrI monolayer was estimated by using the Monte Carlo simulation. Figure~\ref{fig:2}(d) shows the magnetic moment and specific heat of the V atoms as a function of temperature. The calculated \emph{T}$_C$ of the Janus VSBrI monolayer is 83 K, which is higher than that of other magnetoelectric multiferroics, such as VOClBr (43 K)~\cite{A.-J. Phys-2023}, NbSCl$_2$ (25 K)~\cite{Y-J.-2024}, and  VOF$_2$ (15 K)~\cite{H.-Phys.-2020} monolayers.

\begin{figure}[t!hp]
\centerline{\includegraphics[width=0.9\textwidth]{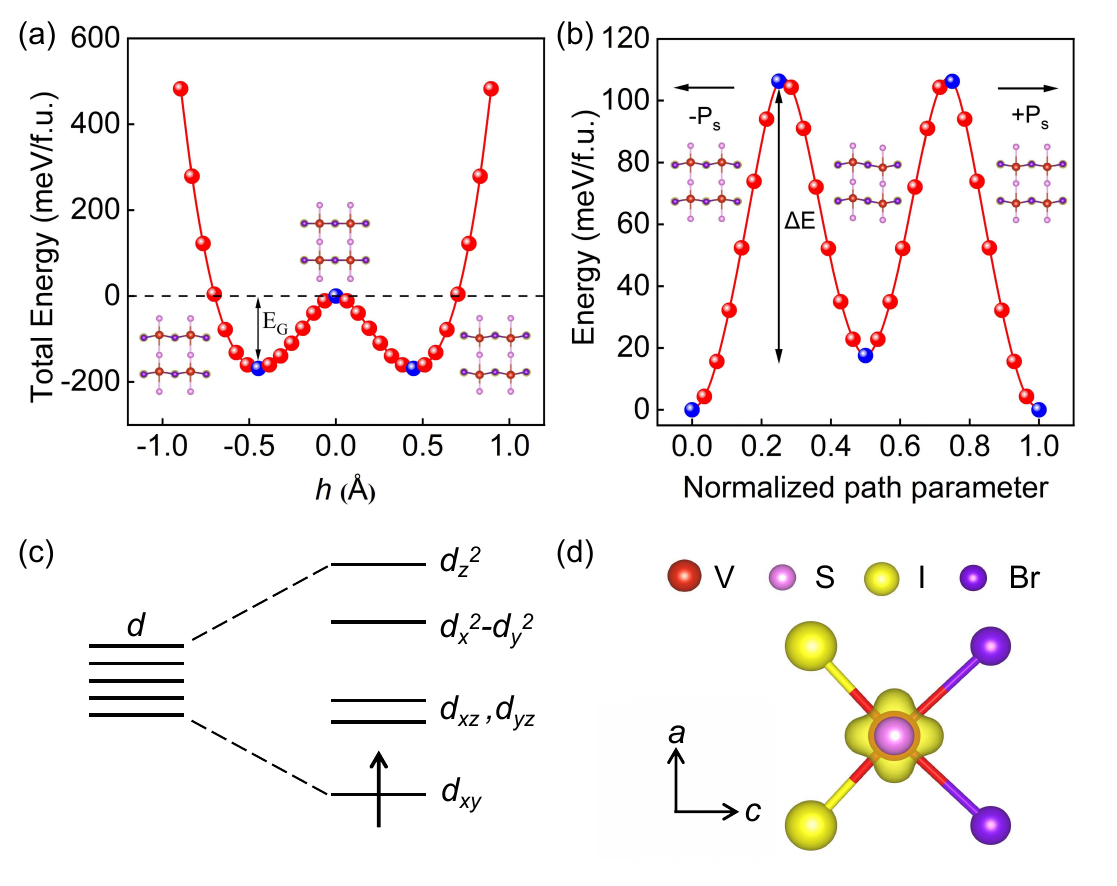}}
\caption{For Janus VSBrI monolayer: In-plane ferroelectric switching pathways via (a) paraelectric (PE) intermediate phase with energy barrier \emph{E}$_G$, and (b) antiferroelectric (AFE) intermediate phase with energy barrier $\Delta$\emph{E}, (c) the orbital splitting due to the crystal field, and (d) the charge density profile of the occupied $d$-orbital.
\label{fig:3}}
\end{figure}

In the Janus VSBrI monolayer, the displacement of the V ions from the middle of the octahedron disrupts inversion symmetry, leading to the generation of a reversible in-plane electric polarization along the \emph{b}-axis. The calculated in-plane spontaneous ferroelectric polarization (\emph{P}$_s$) 1.20 $\times$ $10^{-10}$ C/m, is larger than those of CrNCl (0.60 $\times$ $10^{-10}$ C/m)~\cite{H. -Appl.-2023} and SnTe (0.22 $\times$ $10^{-10}$ C/m)~\cite{S-Rev.-2021} monolayers. Figure~\ref{fig:3}(a) shows the energy versus structural parameter (\emph{h}) for the Janus VSBrI monolayer, which resembles a double-well potential curve. The \emph{h} is defined as the difference between the two V-S bond lengths along the \emph{b}-axis. Here, \emph{h} = 0.45 and -0.45 {\AA} represent the two equivalent stable FE phases with opposite electric polarizations, and \emph{h} = 0 represents the centrosymmetric paraelectric (PE) phase. For the Janus VSBrI monolayer, the calculated energy barrier (\emph{E}$_G$) for ferroelectric switching is 169 meV/f.u, which is higher than that of the parent VSI$_2$ monolayer (140 meV/f.u)~\cite{D.-Appl.-2023}, indicating that its FE phase is relatively more stable. Then, we simulated an in-plane ferroelectric polarization switching pathway through the antiferroelectric (AFE) phase [Fig.~\ref{fig:3}(b)]. The energy barrier ($\Delta$\emph{E}) of the FE-AFE-FE pathway significantly decreases to 88 meV per formula unit. Despite the close energy of the AFE and FE states, the energy barrier between them is significantly larger than the thermal energy at room temperature (25 meV), ensuring that the FE phase remains stable and does not transition to the AFE phase due to thermal vibrations at room temperature. The energy dependence of the three magnetic configurations on polarization was calculated using the ferromagnetic phase as the energy reference [Fig. S3]~\cite{Suppl.Mata.}. The results indicate that the energy differences between the different magnetic states vary significantly with polarization, highlighting the influence of polarization on magnetic properties.  Therefore, it becomes possible for the Janus VSBrI monolayer to exhibit a notable magnetoelectric coupling.

In general, ferromagnetism and ferroelectricity in 2D materials are difficult to coexist due to the magnetism of transition metal $d$ electrons competes with the structural distortions required for ferroelectricity. To preserve a stable magnetic state, $d$ electrons suppress the tendency for off-center ferroelectric distortions. Next, we analyze the reasons for the coexistence of ferromagnetism and ferroelectricity in Janus VSBrI monolayer in detail. Due to the anisotropic octahedral crystal field in the VSBrI monolayer, the cation 3\emph{d} orbitals split into four energy levels: \emph{d$_{z^2}$}, \emph{d$_{x^2-y^2}$}, \emph{d$_{xz}$}/\emph{d$_{yz}$}, and \emph{d$_{xy}$} orbitals, respectively~\cite{H.-Phys. -2019} [Fig.~\ref{fig:3}(c)]. The V$^{4+}$ ion has a 3$d^1$ electronic configuration, with the single unpaired \emph{d} electron occupying the lowest energy \emph{d$_{xy}$} orbital, which is distributed in the plane perpendicular to the V-S chain, as illustrated by the charge density plot in Fig.~\ref{fig:3}(d). Here, although V$^{4+}$ is not \emph{d$_{0}$}, it behaves like \emph{d$_{0}$} along the \emph{z} direction (the \emph{b}-axis) since the \emph{d$_{yz}$} and \emph{d$_{xz}$} orbitals are empty. This unique anisotropic orbital ordering is responsible for the violation of the ``\emph{d$_{0}$} rule"~\cite{N-J. Phys. Chem. B-2000}, resulting in the ferroelectricity, referred to as the ``anisotropic \emph{d$_{1}$} rule"~\cite{N.-Phys. Rev. B-2020}. Therefore, using the anisotropic \emph{d$_{1}$} rule, more 2D magnetoelectric multiferroic materials can be discovered. First, a single-phase ferroelectric structure needs to be identified, and its polarization direction determined. Next, part of the non-metal atoms are replaced with transition metal atoms, ensuring that the transition metal ion has only one electron in the $d$ orbital. Then, based on the crystal field distribution, the occupied orbital (such as \emph{d$_{xy}$},  \emph{d$_{xz}$}, \emph{d$_{yz}$}, or \emph{d$_{x^2-y^2}$}) is determined, and whether its direction aligns with the polarization direction is assessed. This method allows for the preliminary screening of 2D magnetoelectric multiferroic materials.

\begin{figure}[t!hp]
\centerline{\includegraphics[width=1.0\textwidth]{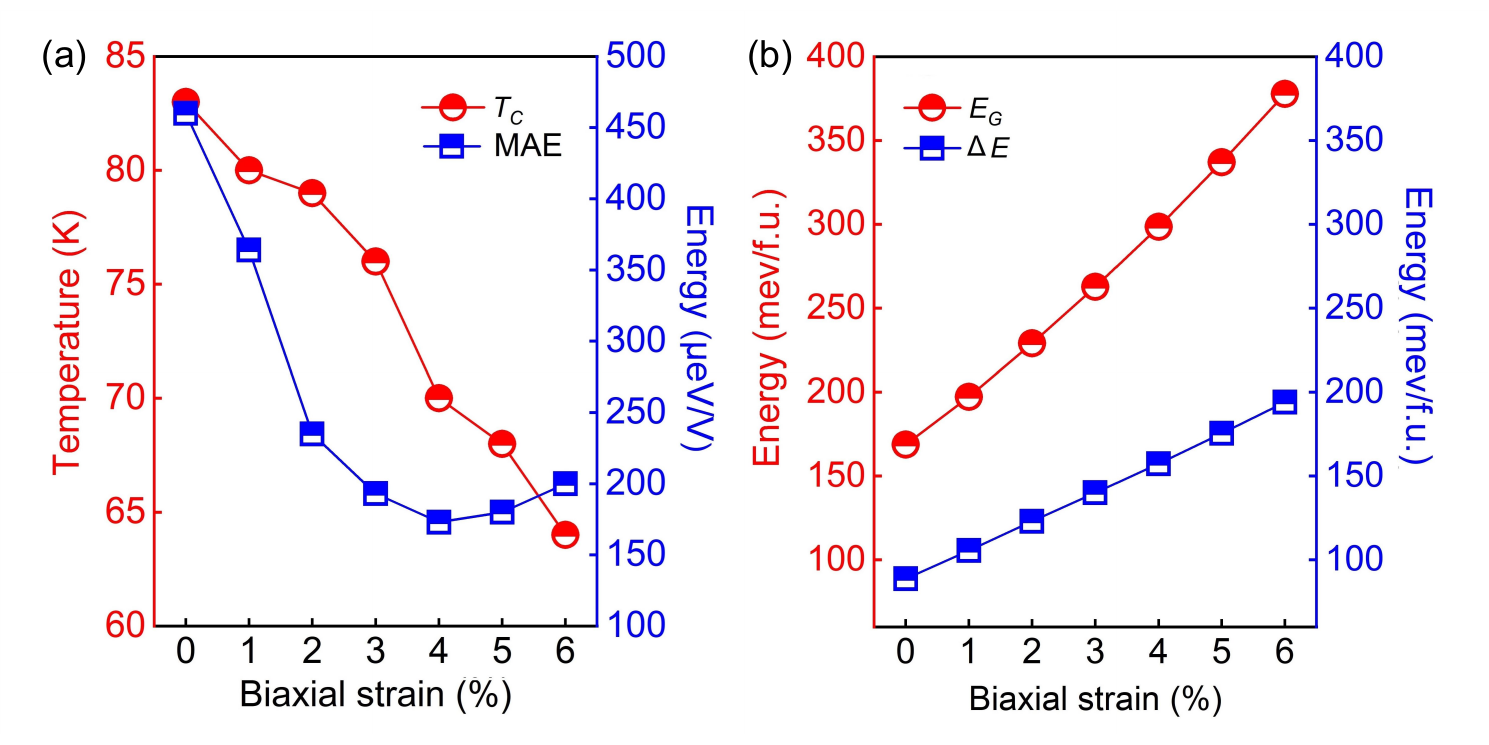}}
\caption{(a) The Curie temperature (\emph{T}$_C$) and magnetic anisotropy energy (MAE), (b) energy barrier (\emph{E$_G$} and $\Delta$\emph{E}) of the VSBrI monolayer under biaxial strain ranging from 0\% - 6\%.
\label{fig:4}}
\end{figure}

Biaxial strain engineering is an effective approach to tuning the properties of 2D materials. We investigated the impact of biaxial tensile strain ranging from 0\% to 6\% on the magnetic, electronic, and ferroelectric properties of the Janus VSBrI monolayer. It exhibits strong ferromagnetic robustness under biaxial strain from 0\% to 6\% [Fig. S4]~\cite{Suppl.Mata.}. Fig.~\ref{fig:4}(a) shows the variation of \emph{T}$_C$ and MAE under biaxial strain ranging from 0\% to 6\%. Despite the gradual decrease in \emph{T}$_C$ with increasing biaxial tensile strain, it remains higher than the temperature of CrI$_3$ (45 K)~\cite{B.-Nature-2017}. Calculations reveal that when a 6\% biaxial tensile strain is applied, the MAE value along the \emph{z}-axis is the lowest, causing the magnetic easy axis to shift from the \emph{x}-axis to the \emph{z}-axis. The Janus VSBrI monolayer maintains its semiconductor behavior, with the spin gap gradually increasing as the tensile strain increases, reaching 0.92 eV at 6\% strain [Fig. S5]~\cite{Suppl.Mata.}. The results show that the polar displacement and erroelectric polarization increase monotonically with the application of biaxial tensile strain [Fig. S6]~\cite{Suppl.Mata.}. The in-plane ferroelectric polarization reaches its maximum value of 1.43 $\times$ $10^{-10}$ C/m under 6\% biaxial tensile strain. The \(E_{G}\) and \(\Delta E\) values increase with the biaxial tensile strain, ranging from 168 to 378 meV/f.u. and 89 to 194 meV/f.u., respectively [Fig.~\ref{fig:4}(b)]. It is known that high-energy barriers can ensure the stability of the ferroelectric phase. By studying the effect of biaxial tensile strain on ferroelectric properties, it was found that the stability of the ferroelectric phase in Janus VSBrI monolayer is further enhanced with increasing biaxial tensile strain.

In noncentrosymmetric materials, the piezoelectric effect leads to the generation of electric dipole moments and electric charge in response to applied mechanical strain or stress. The 2D piezoelectric effect is described by the piezoelectric stress coefficients \emph{e$_{ij}$} and the piezoelectric strain coefficients \emph{d$_{ij}$} (More details are given in the Supplemental Material~\cite{Suppl.Mata.}). The calculated \emph{e$_{11}$} and \emph{e$_{12}$} are 11.10 $\times$ $10^{-10}$ C/m and 3.17 $\times$ $10^{-10}$ C/m, respectively, while the \emph{e$_{31}$} and \emph{e$_{32}$} are 0.40  $\times$ $10^{-10}$ C/m and 0.56 $\times$ $10^{-10}$ C/m, respectively. According to equations (8)-(11) of the Supplemental Material, the calculated in-plane piezoelectric strain coefficients \emph{d$_{11}$} and \emph{d$_{12}$} are 29.01 pm/V and 3.17 pm/V, respectively. The \emph{d$_{11}$} value for the VSBrI monolayer is much larger than that of the well-known piezoelectric material MoS$_2$ (3.73 pm/V)~\cite{L.-ACS-2017}. In contrast to the centrosymmetric VSI$_2$ with only in-plane piezoelectricity, the absence of inversion symmetry in VSBrI results in two distinctive out-of-plane piezoelectric responses characterized by \emph{d$_{31}$} and \emph{d$_{32}$}. The \emph{d$_{31}$} (0.75 pm/V) of the Janus VSBrI monolayer is much larger than that of the Janus group-III chalcogenide monolayers (0.07-0.46 pm/V)~\cite{Y-Appl.-2017}. Interestingly, the \emph{d$_{32}$} is more than twice \emph{d$_{31}$}, with a value of 1.60 pm/V, which is larger than those of many known 2D materials, making it highly desirable for multifunctional piezoelectric devices.

Considering the valence electron is from a 3\emph{d} orbital for Janus VSBrI monolayer, the Hubbard $U$ correction was reassessed using the GGA+$U$ method (\emph{U$_{eff}$} = 0 $\sim$ 2 eV) to re-check its magnetic, ferroelectric, structural, and electronic properties. The results confidently concluded that the magnetic ground state of the Janus VSBrI monolayer remains both FM and FE, unaffected by the choice of the Hubbard $U$ in DFT calculations, and the selected \emph{U$_{eff}$} values have little effect on the structural and electronic properties. (More details are given in the Supplemental Material)~\cite{Suppl.Mata.}.

\busedit{\st{\section{CONCLUSION}}}
In summary, our study predicts that the Janus VSBrI monolayer is a promising 2D intrinsic semiconductor with a band gap of 0.76 eV. This material exhibits notable properties, including ferroelasticity, ferroelectricity, and piezoelectricity. The \emph{T}$_C$ and MAE reach 83 K and 460 $\mu$eV/V, respectively. The spontaneous ferroelectric polarization along the $b$-axis is calculated to be approximately 1.20 $\times$ $10^{-10}$ C/m. In-plane ferroelectric switching pathways involve a polarizable intermediate phase with an energy barrier of 169 meV/f.u. and an antiferroelectric phase with a barrier of 88 meV/f.u. Notably, the VSBrI monolayer displays significant magnetoelectric coupling, attributed to substantial energy differences between various magnetic states influenced by polarization. Furthermore, the stability of the ferroelectric phase is enhanced under increasing biaxial tensile strain. With predicted in-plane piezoelectric coefficients \emph{d$_{11}$} (29.01 pm/V) and out-of-plane \emph{d$_{32}$} (1.60 pm/V) surpassing many existing 2D materials, this monolayer holds great potential for applications in ultrathin piezoelectric devices. Our findings lay the groundwork for advancing multiferroic and piezoelectric materials, which are vital for the development of multifunctional spintronic devices.


\busedit{\st{\section{ACKNOWLEDGEMENT}}}

This work was supported by the Natural Science Foundation of China under Grants (No.22372142), the Innovation Capability Improvement Project of Hebei province (22567605H), the Natural Science Foundation of Hebei Province of China (No. B2021203030), the Science and Technology Project of Hebei Education Department (No. JZX2023020). The numerical calculations in
this paper have been done on the supercomputing system in the High Performance Computing Center of Yanshan University.


\busedit{\st{

\end{document}